\documentclass[a4paper, oneside, twocolumn, notitlepage, 10pt]{extarticle_ecoc}
\usepackage{ecoc}
\usepackage{booktabs}
\usepackage{amsmath,amssymb,amsfonts}
\usepackage{algorithmic}
\usepackage{graphicx}
\usepackage{subfigure}
\usepackage{textcomp}
\usepackage{tikz}
\usepackage{xcolor}
\usepackage{pgfplots}
\usepackage{pgfplotstable}
\usepackage{tikz,pgfplots}
\usetikzlibrary{positioning,shadows,shapes,backgrounds,calc}
\usepackage{pgfplotstable}
\pgfplotsset{compat=1.12} 
\usepackage{subfigure}
\usepackage{upgreek}
\usepackage{tikz}
\usepackage{indentfirst} 
\usepackage{amsfonts,amssymb} 
\usepackage{amsmath}
\usepackage{mathrsfs}
\usepackage{multirow}
\usepackage{pgfplots}
\usepackage{amsmath}
\usepackage{booktabs}

\usepackage{amssymb}
\usepackage{booktabs}
\usepackage{float}
\usepackage{multirow}
\usepackage{array}
\makeatletter

\newcommand{\Rmnum}[1]{\expandafter@slowromancap\romannumeral #1@}
\makeatother
\usetikzlibrary{chains,scopes,positioning,backgrounds,shapes,fit,shadows,calc,arrows.meta}
\usepackage{bm}
\usepackage{tikz,pgfplots}
\usetikzlibrary{calc}
\usepackage{graphicx}
\usepackage{float}
\usepackage{epstopdf}
\usepackage{pgfplotstable}
\usepackage{pgfplots}
\usepackage{subfigure}
\definecolor{myGreen}{rgb}{0.00000,0.49804,0.00000}
\usepackage[utf8]{inputenc}

\usepackage{tikz}
\usetikzlibrary{arrows, calc, fit, math, positioning, shapes.geometric}

\usepackage{balance} 

\definecolor{orange}{rgb}{1,0.7,0}
\definecolor{myDarkGreen}{rgb}{0.00000,0.58824,0.00000}%

\definecolor{myGreen}{rgb}{0.00000,0.49804,0.00000}

\definecolor{blue}{rgb}{0.38, 0.51, 0.71} 
\definecolor{darkblue}{RGB}{17, 42, 60} 
\definecolor{red}{RGB}{175, 49, 39} 

\definecolor{orange}{RGB}{217, 156, 55} 
\definecolor{green}{RGB}{144, 169, 84} 
\definecolor{palegreen}{RGB}{197, 184, 104} 

\definecolor{yellow}{RGB}{250, 199, 100} 
\definecolor{brokenwhite}{RGB}{218, 192, 166} 
\definecolor{brokengrey}{rgb}{0.77, 0.76, 0.82} 

\usetikzlibrary{patterns}
\usepackage{arydshln}
\begin{document}
\selectlanguage{english}    


\title{Experimental Demonstration of  16D Voronoi Constellation  with Two-Level Coding over 50km Four-Core Fiber
}%


\author{
    Can Zhao\textsuperscript{(1)}, Bin Chen\textsuperscript{(1),*}, 
    Jiaqi Cai\textsuperscript{(2)}, Zhiwei Liang\textsuperscript{(1)}, Yi Lei\textsuperscript{(1)}, \\Junjie Xiong\textsuperscript{(3)}, Lin Ma\textsuperscript{(3)}, Daohui Hu\textsuperscript{(2)}, Lin Sun\textsuperscript{(2)}, Gangxiang Shen\textsuperscript{(2)}
}

\maketitle                  


\begin{strip}
    \begin{author_descr}

        \textsuperscript{(1)} School of Computer Science and Information Engineering, Hefei University of Technology, Hefei, China
        *Corresponding author: \textcolor{blue}{\uline{bin.chen@hfut.edu.cn}} 

        \textsuperscript{(2)} School of Electronic and Information Engineering, Soochow University, Suzhou, China

       \textsuperscript{(3)} State Key Laboratory of Advanced Optical Communication Systems and Networks, Shanghai Jiao Tong University, Shanghai, China.

    \end{author_descr}
\end{strip}

\renewcommand\footnotemark{}
\renewcommand\footnoterule{}


\begin{strip}
    \begin{ecoc_abstract}
        A 16-dimensional Voronoi constellation  concatenated  with multilevel coding  is  experimentally demonstrated over a 50~km four-core fiber transmission system. The  proposed  scheme reduces the required launch power by  6~dB and provides a 17~dB larger  operating range  than 16QAM with BICM at the outer HD-FEC BER threshold.
        \textcopyright2024 The Author(s)
    \end{ecoc_abstract}
\end{strip}


\section{Introduction}
The continuously growing capacity demands are driving the data rates of optical fiber transmission systems closer to the Shannon limit of single-mode fiber (SMF) \cite{essiambre2010capacity}. 
To overcome the coming capacity crunch, multi-core fiber (MCF) emerges as a promising space-division multiplexing (SDM) technology for next-generation high-speed optical transmission systems.  
MCF not only provides more parallel channels for achieving  higher capacity, but also supports the joint transmission of multi-dimensional (MD) modulation formats across different cores.
Differently from transmitting independent symbols over each core, such as quadrature amplitude modulation (QAM), the use of MD formats allows to encode data jointly on multiple dimensions or to optimize the distribution of constellation points, bringing higher performance \cite{forney1984efficient,karlsson2016multidimensional}.

Experimental demonstrations have been performed by applying MCF to transmit MD modulation formats, such as twelve-dimensional (12D) \cite{12D2020first} and sixteen-dimensional (16D) modulation formats\cite{rademacher2015experimental}.
However, these MD formats are constrained to low spectral efficiencies (SEs)\cite[Table~I,]{ChenJLT2023},  such as a maximum of 2~bit/2D-sym. Increasing SE and dimensionality will make constellation shaping design difficult, due to the large degree of freedom and the nonconvex MD optimization problem.  

\par
To overcome the bottleneck of SE and complexity limitation for MD formats, MD Voronoi constellation (VC) as an advanced geometric shaping method,  offers high shaping gain and low complexity at high SE by eliminating the need for look-up tables to store constellation points \cite{ALI_low}. VCs have been investigated both in additive white Gaussian noise (AWGN) channel \cite{integertoVC,li2022power} and optical transmission experiments across multiple physical dimensions (wavelength, polarization and time slots) of SMF \cite{Ali2021,Ali_physical}.  By combining MD VC  with multi-level coding (MLC),  the performance of optical transmission system can be further enhanced \cite{lishen2023coded}. 
Recently, we have shown that the joint transmission of 16D VC with MLC outperforms the independent transmission of 16QAM with bit-interleaved coded modulation (BICM) through a four-core MCF via  simulations, which shows up to 0.65~dB signal-to-noise (SNR) gain in AWGN channel and  21\%  reach increase in a four-core optical fiber channel \cite{OL2024}.

In this paper, we experimentally investigate the performance of 16D VC compared to 16QAM in a 50~km four-core  weakly-coupled MCF (WC-MCF) transmission scenario,  achieving a 6~dB minimum launch power reduction with up to 23~dB optical power operating range at the HD-FEC BER threshold. To the best of our knowledge, this is the first experimental demonstration of a 16D VC that spans across a four-core MCF.

\section{MD Vonronoi constellation with MLC }
\begin{figure*}[!tb]
\vspace{-0em}
\centering
\includegraphics[width=1.0\textwidth]{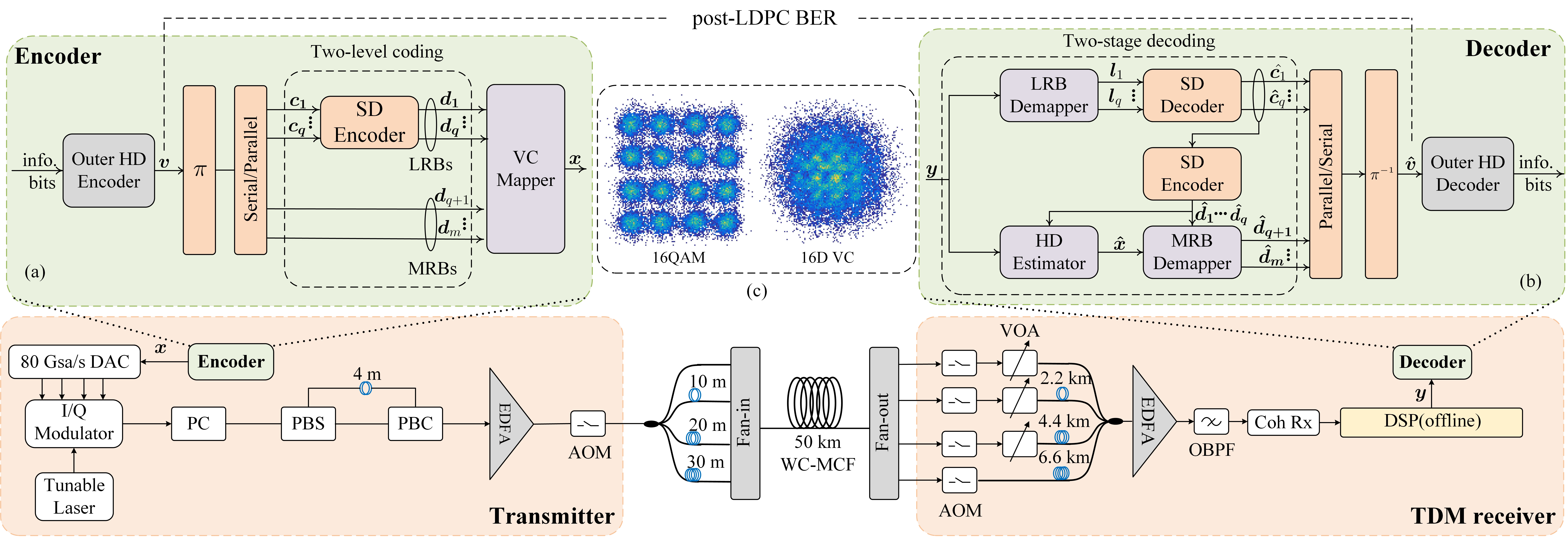}
\vspace{-0.7em}
\caption{Experimental setup. Inset (\textbf{a}) shows the encoder of VC with MLC scheme. Inset (\textbf{b}) shows the decoder of VC with MLC scheme. Inset (\textbf{c}) shows the 2D projection of the constellations after 50~km transmission at the optimal launch power.}
\label{ep}
\end{figure*}

VCs are structured MD lattice-based constellations comprising a coding lattice and a shaping lattice \cite{forney1984efficient}. The coding lattice dictates the arrangement of the constellation points, resulting in coding gain, while the shaping lattice determines the shape of the constellation boundary, yielding shaping gain. The combination of different shaping and coding lattices in MD space can result in a higher and more granular set of possible SEs, providing greater flexibility in selecting data rates  \cite{ALI_low}.

MLC as an advanced coded modulation (CM) scheme, provides a better complexity-performance trade-off than BICM.  This is achieved by selectively protecting the least reliable bits (LRBs) of a mapper via an inner FEC code, instead of uniformly protecting all bits as in BICM. The bit error rate (BER) performance between BICM for VC and MLC for VC has already been compared \cite{lishen2023coded}, where MLC for VC exhibits better performance. The encoder and decoder of a two-level MLC scheme for MD VC are shown in Fig.~\ref{ep} inset (a) and inset (b).

For the encoder (information bits to VC symbols $\bm{x}$), an outer hard-decision (HD)  code with 0.9373 coding rate is considered \cite{smith2011staircase}.
Information bits are first encoded into $\bm{v}$ by an outer encoder, and then are passed through an interleaver $\pi$ to prevent burst errors. 
The serial bits after interleaver are partitioned into $q$ parallel bit streams $\bm{c}_i$ ($i$ = 1, \ldots, $q$)  and $m - q$ parallel bit streams $\bm{d}_{q+1}$, \ldots, $\bm{d}_{m}$. In the two-level coding, the $q$ parallel bit streams are encoded by an inner low-density parity-check (LDPC) encoder and are then used as the LSBs of the MD-VC symbols, while the $m - q$ parallel bit streams are directly used as the most reliable bits (MRBs). The mapping of LSBs and MRBs to VC points is accomplished through the VC mapper \cite [Alg.~3] {lishen2023coded,integertoVC}.

\begin{table}[!tb]
 \caption{Parameters used for BICM-16QAM  and MLC-16D VC}
 \footnotesize
    \centering
    \def\arraystretch{1}
\begin{tabular}{c|c|c}
\hline\hline
{\bf Constellation} & {\bf 16QAM} & {\bf 16D-VC}\\
\hline
CM & BICM & MLC\\
\hline
SE  (bit/2D-sym) & 4 & 4.5\\
\hline
LDPC code rate (inner) & 8/9 & 1/2\\
\hline
Coded bit levels/$m$ & 4/4 & 16/36\\
\hline
Information rate (bit/2D-sym) & 3.56 & 3.5\\
\hline\hline
\end{tabular}
    \label{tab:para}
    \vspace{-2em}
\end{table}

For the decoder (received symbols $\bm{y}$ to recovered bits), a two-stage decoding process is performed. In the first stage, the LRB demapper calculates the log-likelihood ratios (LLRs) of the LRBs ${\bm{d}}_i$ \cite[Eq.~(34),]{lishen2023coded}, which are then decoded to obtained the estimated bits $\bm{\hat{c}_i}$ ($i$ = 1, \ldots, $q$).
In the second stage, the inner soft-decision (SD) encoder encodes  $\bm{\hat{c}_i}$ obtained in the first stage into $\bm{\hat{d}}_i$, and the estimated transmitted symbols $\hat{\bm{x}}$ are obtained by applying the closest point algorithm in the HD estimator \cite[Eq.~(37),]{lishen2023coded}. Based on $\hat{\bm{x}}$ and $\bm{\hat{d}}_i$, the estimated MRBs ($\bm{\hat{d}}_{q+1},\ldots,\bm{\hat{d}}_{m}$) are finally obtained by the MRB  demapper. Detailed demapping algorithms are discussed in   \cite [Alg.~4,] {lishen2023coded,integertoVC}. Subsequently, the estimated LRBs and MRBs go through parallel/serial conversion and deinterleaver to obtain $\bm{\hat{v}}$.

\begin{figure}[!tb]
\vspace{-0.3em}
\centering
\includegraphics[width=0.48\textwidth]{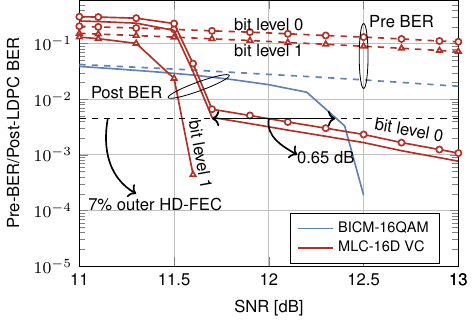}
\vspace{-1.5em}
\caption{Simulation results of  BER  for two coded modulation schemes in AWGN channel. The triangle markers  and circle markers denote bit level 0 and bit level 1, respectively.  Dashed: pre-LDPC BER; Solid: post-LDPC BER.}
\label{simulation}
\vspace{-0.5em}
\end{figure}

AWGN simulations are first performed to compare the performance of 16D VC and 16QAM. The simulation parameters are given in Table~\ref{tab:para}.
Fig.~\ref{simulation} shows the pre- and post-LDPC BER performance of16QAM with BICM and VC with MLC in AWGN channel around the overall information rate of 3.5~bit/2D-sym. It can be seen that the post-LDPC BER of both schemes initially exhibit same trends, i.e., keeping flat first and then decreasing sharply when beyond an SNR threshold. However, VC with MLC has a high error floor at the outer BER threshold. The reason is that BICM protects all bits, leading to the sharply decrease of  BER with the help of LDPC. 
For MLC, bits are paritioned into two reliability levels that bits in level 1 are protected by LDPC, while bits in level 0 are unprotected. As LDPC corrects the errors in level 1, the post BER of both bit levels exhibit a sharp decrease. However, once bit level~1 achieves near error-free performance, the performance of bit level 0 depends on the knowledge of its fully correct information, resulting in a high error floor at the HD-FEC threshod for MLC. Nevertheless, MLC still offers 0.65~dB  SNR gain compared to QAM with BICM at the HD-FEC BER threshold.

\section{Experimental setup}
The experimental setup for the four-core MCF transmissions of 16D VC with MLC is illustrated in Fig.~\ref{ep}. 
A dual-polarization 20~GBaud 16D VC optical signal is modulated by a tunable laser at 1550~nm using an 80~GSa/s digital-to-analog convert (DAC), a dual-polarization IQ-Modulator and a polarization controller (PC).
To ensure the independence of the two optical polarization symbols, about 407 relative symbols are delayed by a 4~m delay fiber which connects the polarization beam splitter (PBS) and polarization beam combiner (PBC). The signal is amplified by an erbium-doped fiber amplifier (EDFA), split into 4 paths with an acousto-optic modulator (AOM) to control the loading time, decorrelated by 0, 1001 (10~m), 2002 (20~m) and 3003 (30~m) symbols, and fed into the four cores of a 50~km  four-core MCF, respectively. 

\begin{figure}[!tb]
 \hspace{-0.5em}
\centering
\includegraphics[width=0.5\textwidth]{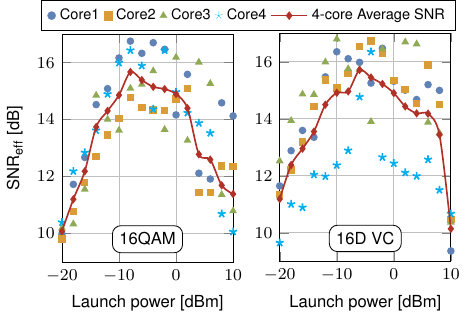}
\caption{Effective SNR as a function of the launch power over 50~km transmission in WC-MCF.}
\label{Exp_SNR}
\end{figure}

At the receiver side, variable optical attenuators (VOAs) are combined with AOMs to control the output power of the four cores. The coupled four-core signals  are transmitted in series into a time-division-multiplexing (TDM) receiver\cite{TDM} with 11~$\upmu$s (2.2~km), 22~$\upmu$s (4.4~km) and 33~$\upmu$s (6.6~km) delays to reduce the amount of required receivers. Then the four paths signals are combined by a coupler, amplified, filtered by an optical bandpass filter (OBPF), and digitized by a coherent receiver consisting of a local oscillator (LO) and a 128-GSa/s analog-to-digital converter (ADC). Offline digital signal processing (DSP)
includes serial/parallel (S/P) conversion, resampling, clock and data recovery (CDR), frequency-offset compensation (FOC), chromatic dispersion (CD) compensation, decision-directed least mean square (DD-LMS), and carrier phase estimation (CPE).
To calculate BER, the received symbols are transmitted to the decoder to implement two-stage decoding.

\section{Experimental Results}
Fig.~\ref{Exp_SNR} shows the effective SNR (after fiber propagation and receiver DSP) of 16QAM and 16D VC formats for each core over a 50~km WC-MCF. As the average SNR (red line) shown, the SNR of each core is centrally distributed near the red line when transmitting the 16QAM formats (left figure) in the linear regime. However, the differences between cores become larger as the launch power increases. Conversely,  for the 16D VC format shown in the right figure, it can be observed that the SNR distribution appears more dispersed than QAM in any launch power. 

\begin{figure}[!tb]
 \hspace{-0.5em}
\centering
\includegraphics[width=0.5\textwidth]{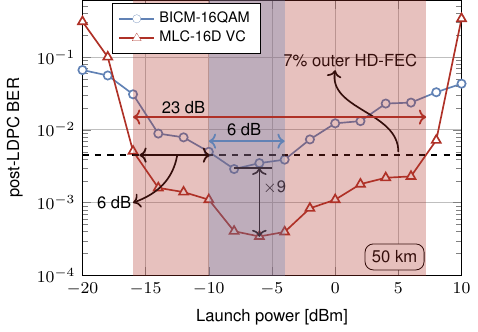}
\vspace{-0.5em}
\caption{Post-LDPC BER as a function of launch powers via 4-core transmission for 16QAM and 16D-VC with  the net  information rate  around 530~Gbps.}
\label{Exp_BER}
\end{figure}

Fig.~\ref{Exp_BER} shows the post-LDPC BER as a function of different launch powers for VC with MLC in comparison to QAM with BICM, with a net information rate around 530~Gbps while considering a 6.7\% overhead staircase code as the outer code. In the linear regime, QAM initially outperforms VC. However, as the launch power increases, VC performs better than QAM. The reason for this is  that the post-LDPC BER of MLC-16D VC is limited by error correction capability, and a low LDPC rate leads to more errors at low SNRs. Nonetheless, the advantage of MLC improves more than BICM at higher SNRs. 
The  minimum launch power required at the outer BER threshold 
is reduced approximately 6~dB. The red and blue shades show the operating regions of VC and QAM, respectively. It can be observed that VC exhibits a 23~dB working region, while QAM exhibits only a 6~dB operating range, with over 17~dB  larger launch  power dynamic range   at the HD-FEC BER threshold. 
When comparing the best performance of VC with QAM, the post-LDPC BER improvements are approximately  a factor of 9. 
These gains come from both the shaping gain provided by VC and the coding gain provided by MLC.
Though the SNR variation among the four cores of VC are larger than QAM in Fig.~\ref{Exp_SNR}, VC can still exhibit a better performance. 

\section{Conclusion}
We experimentally demonstrate the transmission of a 16D VC concatenated with a two-level MLC over 50~km of a four-core fiber.
A launch power gain of 6~dB over QAM with BICM is shown, and with up to 17~dB operating range gains due to the joint decoding of MD VC. 
Experimental results are in good agreement with simulation of our previous work, thus confirming the potential benefits of employing MD VC in MCF transmission system. 

\section{Acknowledgements}
The authors are grateful to Shen Li from Université Laval, Canada for 
for fruitful discussions on Voronoi Constellation. 
This work is supported by the National Natural Science Foundation of China (62171175, 62001151), and State Key Laboratory of Advanced Optical Communication Systems and Networks, Shanghai Jiao Tong University, China.

\defbibnote{myprenote}{%
}
\printbibliography[prenote=myprenote]
\vspace{-4mm}

\end{document}